# Cryogenic scanning photocurrent spectroscopy for materials responses to structured optical fields

**Authors**

Duxing Hao[1], Chun-I Lu[1], Ziqi Sun[1),a)], Yu-Chen Chang[2], Wen-Hao Chang[1,2], Ye-Ru Chen[2], Akiyoshi Park[1], Beining Rao[1),a)], Siyuan Qiu[1),b)], Yann-Wen Lan[2),c)], Ting-Hua Lu[2),c)] and Nai-Chang Yeh[1,2,3),c)]

**Affiliations**

[1)] Department of Physics, California Institute of Technology, Pasadena, California, 91125, USA

[2)] Department of Physics, National Taiwan Normal University, Taipei, 116059, Taiwan

[3)] Kavli Nanoscience Institute, California Institute of Technology, Pasadena, California, 91125, USA

[a)] This work was conducted during summer research at California Institute of Technology while being an enrolled undergraduate student at Zhiyuan College, Shanghai Jiao Tong University, Shanghai, 200240, P. R. China.

[b)] This work was conducted during summer research at California Institute of Technology while being an enrolled undergraduate student at Department of Physics, Columbia University, New York, NY, 10027, USA.

[c)] Authors to whom correspondence should be addressed: ywlan@ntnu.edu.tw, thlu@ntnu.edu.tw and ncyeh@caltech.edu

**Abstract**

Circular dichroism spectroscopy is known to provide important insights into the interplay of different degrees of freedom in quantum materials, and yet spectroscopic study of the optoelectronic responses of quantum materials to structured optical fields, such as light with finite spin and orbital angular momentum, has not yet been widely explored, particularly at cryogenic temperature. Here we demonstrate the design and application of a novel instrument that integrates scanning spectroscopic photocurrent measurements with structured light of controlled spin and orbital angular momentum. For structured photons with wavelengths between 500 nm to 700 nm, this instrument can perform spatially resolved photocurrent measurements of two-dimensional materials or thin crystals under magnetic fields up to ±14 Tesla, at temperatures from 300 K down to 3 K, with either spin angular momentum $\pm \hbar$ or orbital angular momentum $\pm \ell \hbar$ (where $\ell = 1, 2, 3\ldots$ is the topological charge), and over a $(35 \times 25)$ $\mu m^2$ area with ~ 1 μm spatial resolution. These capabilities of the instrument are exemplified by magneto-photocurrent spectroscopic measurements of monolayer 2H-$MoS_2$ field-effect transistors, which not only reveal the excitonic spectra but also demonstrate monotonically increasing photocurrents with increasing $|\ell|$ as well as excitonic Zeeman splitting and an enhanced Landé $g$-factor due to the enhanced formation of intervalley dark excitons under magnetic field. These studies thus demonstrate the versatility of the scanning photocurrent spectrometry for investigating excitonic physics, optical selection rules, and optoelectronic responses of novel quantum materials and engineered quantum devices to structured light.

## I. Introduction

Novel optoelectronic properties of materials, such as the light emitting mechanisms and light-controlled matter responses, have attracted intense research interest in the past two decades.[1,2] Two notable ways of engineering light-matter interactions focus on the characteristics of the incident light and the optoelectronic properties of the material. Besides varying the energy and polarization of the photon, the former approach may further include the spin and orbital degrees of freedom. Circularly polarized light that carries a spin angular momentum (SAM) of $\pm\hbar$ has been exploited to achieve selective excitations in materials, especially those without lattice inversion symmetry. On the other hand, twisted light that carries an orbital angular momentum (OAM) of $\pm \ell\hbar$ (where $\ell$ is an integer that represents the topological charge of photons) is believed to have a wide range of applications due to its full utilization of the vector character of the electromagnetic nature of light,[3] which may be employed in such areas as optical tweezers,[4] conventional information transfer,[5] and emerging quantum information technology.[6]

Transition metal dichalcogenides (TMDs) of the 2H phase, 2H-MX$_2$ (M = transition metals, X= S, Se and Te), is a class of two-dimensional (2D) semiconducting quantum materials whose optoelectronic responses may be tuned via strain,[7,8] defect,[9,10] and meta-surface engineering.[11–13] The absence of lattice inversion symmetry and the presence of spin-orbital coupling (SOC) in 2H-MX$_2$ further enriches the fine-tuning handles,[14] allowing for unique chirality-controlled valley-specific excitation and transport.[15] Among the 2H-TMDs, monolayer molybdenum disulfide (ML-2H-MoS$_2$, also known as 1H-MoS$_2$) possesses a direct bandgap of 1.8 eV and SOC-induced valence band splitting of ~150 meV,[16–18] leading to chiral-selected valley-polarized photocurrent[19] and circular polarization-maintained photoluminescence.[20,21] Moreover, under the excitation of *twisted light*, which refers to light with finite orbital angular momentum (OAM) $\pm \ell\hbar$ (where $\ell$ = 1, 2, 3…), enhancement in the photovoltaic effect[22] and intervalley transitions[23] with increasing $\ell$ has been demonstrated at the excitonic resonant energy.

Photocurrent spectroscopy (PCS), a powerful experimental tool for studying the optoelectronic responses of semiconducting materials, measures light-induced photocurrents as a function of the photon energy.[24] This approach is capable of probing the complex excitonic states of semiconductors[25,26] and further identifying dark excitons associated with forbidden optical transitions[27] when the PCS measurements are combined with electrical gating and an external magnetic field. Here we report the development of a scanning PCS system with spatially and SAM/OAM-resolved capabilities and further demonstrate the application of this system to characterizing the optoelectronic responses of ML-MoS$_2$ to SAM/OAM light at cryogenic temperatures and under finite magnetic fields. Our studies reveal a strongly enhanced Landé *g*- factor in ML-MoS$_2$ due to the carrier density-dependent exchange interaction and enhanced formation of intervalley excitons in the presence of magnetic fields, as well as monotonically increasing photocurrents with increasing $|\ell|$ due to the enhanced formation of Rydberg and dark excitons by the OAM light. These findings thus demonstrate the versatility of our scanning PCS system for investigating the excitonic physics, optical selection rules, as well as the optoelectronic responses of novel quantum materials and quantum devices to structured light.

## II. Experimental setup

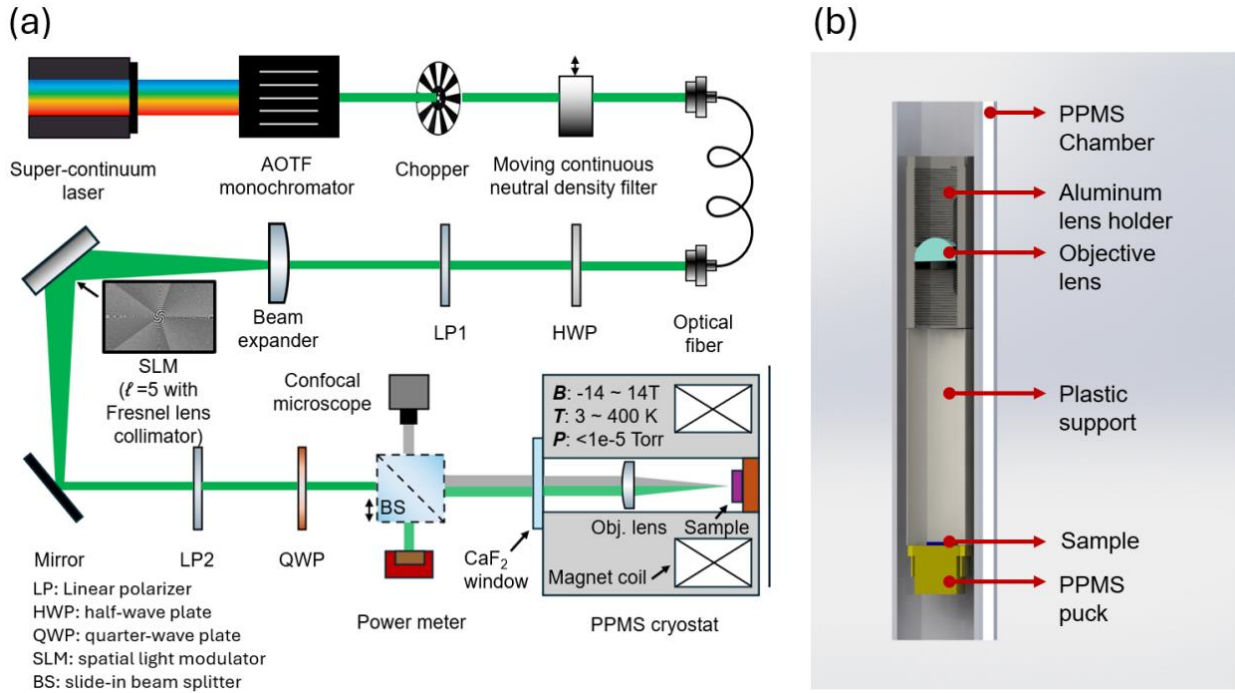

**FIG. 1.** Experimental setup for variable temperature (3 – 400 K) photocurrent spectroscopy (PCS). (a) Schematic layout of the SAM/OAM PCS system. (b) Schematic cross-sectional view of the objective lens holder and support inside the cryostat chamber.

The PCS instrument comprises three major parts: 1) a super continuum laser-based photon generation system, 2) an OAM and SAM generation module with a confocal microscope, and 3) a cryostat with electronic detection and magnetic field control [Fig. 1(a)]. The super-continuum *superK Extreme* laser (NKT photonics) outputs a 4-Watt white light which is filtered using *superK Select* (NKT photonics, VIS 1x). The combined photon generation setup outputs linearly polarized light of wavelengths ranging from 430 nm to 700 nm and a factory-characterized linewidth of 0.5 nm – 1.85 nm, which corresponds to an energy resolution of ~ 3 meV. A mechanical chopper (Thorlabs, MC2000B and MC1F10HP, 20 – 1000 Hz) sets up the base frequency for photon illumination and photocurrent detection using a lock-in technique. The power output was then adjusted by a moving continuous neutral density (cND) filter (Thorlabs, NDL-25C-2, optical density range of 0.04 to 2) carried by a step-motor-controlled translation stage with position repeatability < 3 μm (Zaber-motion, XLSQ600B). Further power adjustment may be conveniently achieved by including additional higher OD neutral density filters in the optical path.

A fiber delivery system (NKT photonics, FD7) with a customized XYZR mount is used to collect the power-adjusted light and to provide a mechanically flexible delivery of light to the customized OAM module. The customized OAM module includes a set of half-wave plate, a linear polarizer, and a beam expander, delivering expanded light onto the spatial light modulator (SLM) with a linear polarization aligned with the extortionary axis of the liquid crystal-based panel of the SLM (Southport, JadeDot-LD). The SLM uses an algorithm-computed 8-bit (1920 pixel × 1200 pixel) pattern that not only generates the OAM pattern for each wavelength but is also capable of x-y beam steering while providing focus adjustment with the Fresnel lens collimator algorithm [Fig. 1(a)]. The OAM module also includes a confocal microscope setup that could be engaged via a slide-in beam splitter. This handy slide-in design helps locate the sample and check the beam quality, which can be removed during measurement to avoid complex wavelength-dependent polarization inconsistency. After the OAM generation, a linear polarizer (Thorlabs, WP25M-UB) and a rotary-mounted (Thorlabs, K10CR2/M) quarter-wave plate (Thorlabs, SAQWP05M-700) are used to generate circular polarization.

The instrument operates in either a room-temperature configuration or a cryostat configuration. The room-temperature setup consists of an achromatic objective lens, a sample holder assembly with the PPMS Userbridge for electronic measurement and a XYZR stage for sample-side positioning. The cryogenic setup is built on a commercial PPMS (Physical Property Measurement System, Quantum Design, Dynacool). A $CaF_2$ optical window and a customized objective lens holder [Fig. 1(b)] is used to achieve collimated SAM/OAM light spot on the sample while conducting concurrent electrical transport measurements.

The instrument achieves a stable temperature of 3 K with 12 mW redundant cooling power under a magnetic field of ±14 T, or 2 K with 4 mW redundant cooling power. To prevent exposure to the stray field up to 10 mT at $|B|$ = 14T, the laser system and the OAM module were enclosed in their customized aluminum shields. The maximum spectroscopic-uniform power output between 500 nm and 700 nm was 20 μW ± 3% with a spot radius of 2.5 μm for zero OAM in the linearly polarized condition, while being chopped at 311 Hz with 50% duty cycle. The wavelength coverage may be extended to the full 430 nm – 700 nm spectral range at the expense of reduced uniformity in the spectral power output due to the sharply decreasing spectral power density of the laser below 500 nm. The power delivered by light to the sample was measured with the room-temperature configuration by placing a power meter (Thorlabs, PM400 and S120C) at the sample location. Further tests revealed that there was no discernible difference in the power reading whether the power meter was placed at the sample position of the room temperature configuration or at a distance comparable to the sample position in the cryostat configuration. Therefore, detailed wavelength-dependent power calibrations to be described below were carried out in the room temperature configuration. We further note that the uniformity of the spectroscopic power output was limited by the long-term stability of the laser (±3% in 2 hours) and the measurement uncertainty of the photodiode head of the power meter (±3%).

Overall, the instrument overcomes several critical challenges: 1) Achieving 500 nm – 700 nm spectroscopic generation of OAM light using a SLM. 2) Integrating PCS with a commercial cryostat setup within a limited space and a long working distance. 3) Scanning a high-quality light spot on micron-scale ML-TMD field-effect transistors (FETs). 4) Performing optical and back-gate tunable electronic transport measurements concurrently under magnetic field up to ±14 T and temperatures down to 3 K.

## III. Evaluation

### A. Spectroscopic orbital angular momentum generation

One of the key challenges for the PCS instrument is that the commercial SLM is usually calibrated at one specified wavelength. To achieve achromatic modulations from the SLM, we need to produce a wavelength-dependent linear mapping between 8-bit greyscale level (*GS*, integer from 0 to 255 as the input of SLM for each pixel) and optical phase modulation 0 to $2\pi$. As shown in Fig. 2(a), the calibration setup consists of two orthogonally polarized linear polarizers placed before and after the SLM, whose extraordinary axis is pointing out from the paper.[28] Assuming that the spatially-uniform phase modulation of the SLM is only on its extraordinary axis and using Jone's matrix $\begin{bmatrix} e^{-i\varphi(GS)} & 0 \\ 0 & 1 \end{bmatrix}$, the measured intensity at the power meter ($I_{out}$) is $I_{out} = I_{in}(1 - \cos\varphi(GS))/4$, where the $\varphi(GS)$ is the phase modulation applied by the SLM at a greyscale level *GS*.

The manufacture's single wavelength hardware calibration procedure was performed first at 700 nm, which resulted in a linear map between the *GS* input 0-255 and the phase depths of 0-2.2 $\pi$ with a phase offset comparing to expected $I_{out}$ described above [Fig. 2(b)]. After the hardware calibration, we measured wavelength dependence between 500-700 nm [Fig. 2(c)] and found two issues that need further calibration: Firstly, roughly 700/$\lambda$ periods were observed for shorter wavelengths, where the *GS* remapping was needed for each wavelength to cut off those excessive periods. Secondly, the observed wavelength-dependent phase offset $\varphi_0(\lambda)$, shown as horizontal shifts in Fig. 2(c), required additional corrections. The above two issues were corrected via a customized Python-based script for the instrument, which first normalized and fitted the measured intensity *I*-vs.-*GS* data to a set of $I_{normalized}(\lambda) = A(\lambda) \sin(GS/T(\lambda) + \varphi_0(\lambda))$ functions, with $A(\lambda)$ and $T(\lambda)$ being the fitted amplitude and period at a wavelength $\lambda$, respectively. Based on the fitted result, a $GS_{start}(\lambda)$ was chosen where the phase modulation of the SLM was 0 and the $GS_{end}(\lambda)$ was calculated as $GS_{end}(\lambda) = GS_{start}(\lambda) + T(\lambda)$. Thus, the greyscale was remapped via the formula: $GS_{remapped}(\lambda) = \text{round}\left(\frac{GS}{255} * (GS_{end}(\lambda) - GS_{start}(\lambda)) + GS_{start}(\lambda)\right)$ such that for the whole range of GS from 0 to 255, the SLM produced a phase modulation linearly from 0 to $2\pi$ for all measured wavelengths. The calibration ensured the input GS stayed in the integer norm and will cause negligible (< 1%) accumulated phase error. In our experiment, the measured wavelength was 500-700 nm with 1 nm step size and the sub-nm correction was achieved by interpolating the experimentally extracted $GS_{start}(\lambda)$ and $GS_{end}(\lambda)$ values using the Pchip interpolator algorithm.[29] The corrected results showed nearly one exact period with negligible phase offset in Fig. 2(d). After carrying out both hardware and software calibrations, signature donut-shape spots for the OAM light of different $\ell$ values with phase singularity at the center of each light spot was achieved [Fig. 2(e)] on a standard silicon chip mounted at the sample end.

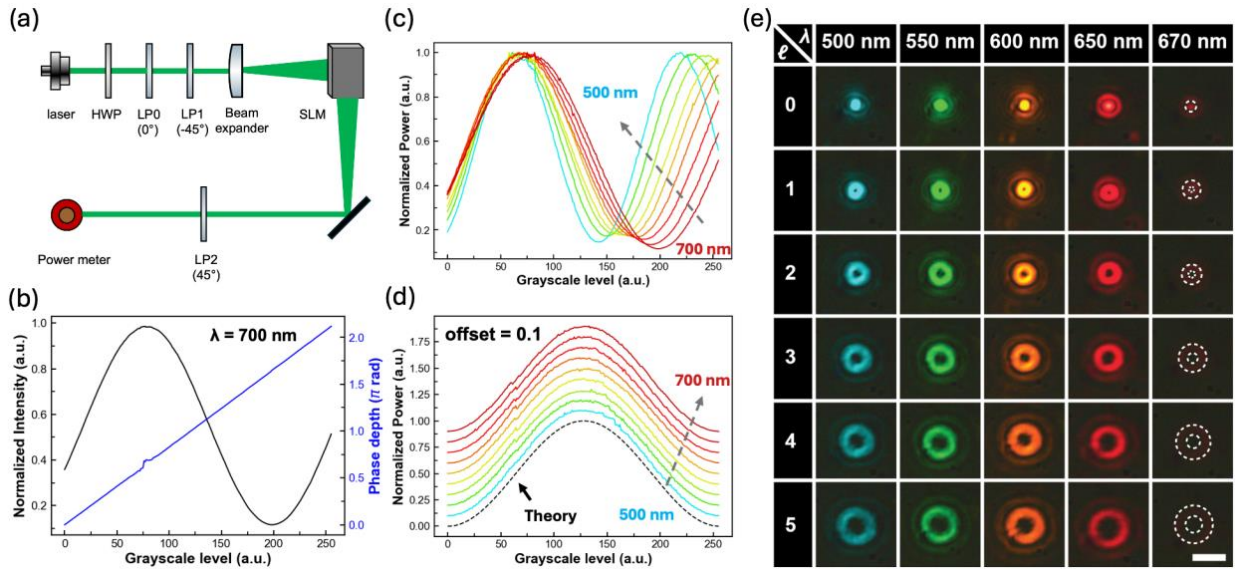

**FIG. 2.** Generation of OAM-light spectroscopy: (a) Schematics of the optical experiment layout. (b) Measured power (left axis) and indicated phase depth (right axis) versus the grayscale level (GS) after hardware calibration. [(c),(d)] Wavelength dependence of normalized light power *vs.* applied GS before (c) and after (d) software calibration. The black dashed line showed the theoretical expectation, *i.e.*, $I_{out} \propto [1 - \cos\varphi(GS)]$. (e) Table of the generated OAM light spots for different wavelengths and $\ell$ values after both hardware and software calibrations, captured by a confocal microscope of the setup at the sample end on a standard Si/SiO$_2$ chip. The white scale bar represents 10 μm. Due to the internal CCD filter, the intensity of the spot image drops sharply for $\lambda > 650$ nm. Therefore, thin white dashed lines were added to outline the inside and outside rims of the donut-shape spots in the $\lambda = 670$ nm column for visual clarity.

## B. Spectral power uniformity

To ensure that the measured spectroscopic photocurrents are purely from the optoelectronic responses of the sample, spectroscopic dependences on the laser power due to optical components of the instrument, such as wave plates and polarizers, must be eliminated. Therefore, a spectral power calibration for the instrument is needed. The instrument employed a moving continuous neutral density (cND) filter to adjust the system power output, and a power meter placed at the sample position was used to measure the power output [Fig. 3(a)]. The power $P = f(x, \lambda)$ measured at spot position $x$ and wavelength $\lambda$ was then interpolated and used as a calibration file such that when a desired power $P_1$ at wavelength $\lambda_1$ was needed, the corresponding position $x_1$ of the cND filter could be queried as $x_1 = f^{-1}(P_1, \lambda_1)$ via an integrated computer algorithm. With such corrections, the instrument with a constant spectroscopic power output (power fluctuation $< \pm 3\%$) was achieved [Fig. 3(b) colored dots], in stark contrast to the highly wavelength dependent output power without calibration [Fig. 3(b) black dots] due to the spectral power density characteristics of the super-continuum laser. The SAM and OAM spectral power uniformity was also tested for various wavelengths as shown in Figs. 3(c) and 3(d), respectively. These results indicated that a consistent spectral power uniformity within $\pm 3\%$ was achieved under varying polarizations, OAM, and power setpoints, thus validating the suitability of our instrument for spectroscopic photocurrent measurements.

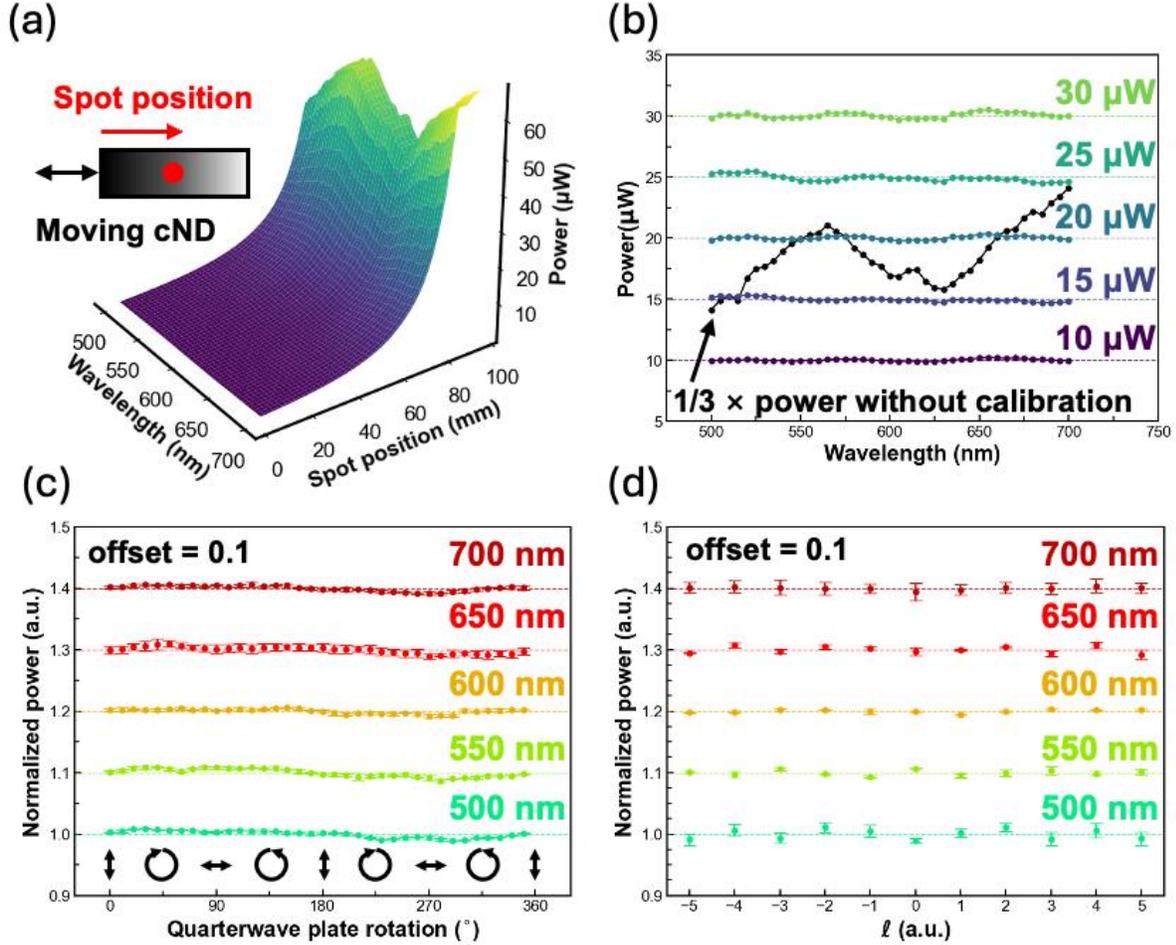

**FIG. 3.** Spectral power uniformity measurement and performance testing. (a) Sample-side power as a function of wavelength and the spot position passing through a moving continuous neutral density (cND) filter. The inset illustrates the spot position relative to the moving cND filter. (b) Measured laser raw power output (black dots) before system spectral power output calibration and measured constant spectral output (colored dots) at 10, 15, 20, 25 and 30 µW setpoints after calibration. (c) Measured polarization-dependent power normalized by a setpoint of 20 µW for wavelength between 500 nm to 700 nm. (d) Measured $\ell$-dependent power normalized by a setpoint of 20 µW for wavelength between 500 nm to 700 nm. All measurements were performed at the sample position in the room-temperature configuration.

### IV. Implementing the scanning SAM/OAM-PCS

The performance of the scanning SAM/OAM-PCS system was evaluated on ML-MoS$_2$ FETs. The ML-MoS$_2$ single crystal samples used for the FETs were synthesized on sapphire substrates by chemical vapor deposition (CVD), and the device fabrication processes were similar to those reported before for electrical transport studies[30,31] except that an additional 5 nm-thick h-BN layer between MoS2 and SiO2 substrate [Fig. 4(a)] were included for the FETs devices used in the PCS studies to eliminate magnetic field-induced polar order[30] in ML-MoS2 on SiO2 substrate at cryogenic temperatures.

For measurements of the photocurrent, a 100 kΩ (± 0.01%) precision resistor served as both a protection resistor and an AC voltage measurement target, as illustrated in Fig. 4(a). A DC bias voltage $V_{DS}$ across the sample and the precision resistor series was applied to achieve a channel current of 10 nA/μm. Unless specified, all photocurrent measurements were taken with the cryogenic configuration at 3 K. The photocurrent $I_{pc}$ was measured with a lock-in amplifier across the precision 100 kΩ resistor in series with the sample at a chopper frequency of 311 Hz. The backgate voltage was swept to a negative voltage of –20 V before obtaining each spectrum in order to remove the residual carriers.[32]

In the following, we illustrate three applications of the PCS system to deriving various optoelectronic properties of ML-MoS2, which include: 1) SAM-resolved PCS measurements in magnetic fields for determining the effective Landé g-factor; 2) scanning PCS measurements for spatially resolved excitonic spectra; and 3) OAM-resolved PCS studies of enhanced optoelectronic responses induced by OAM light.

### A. SAM-resolved photocurrent spectroscopy

Photocurrent spectra taken on a ML-MoS2 FET under both the right circularly polarized (RCP) light and left circularly polarized light (LCP) for applied magnetic field between $B = \pm 14$ T perpendicular to ML-MoS2 were shown in Fig. 4(b). All spectra taken in different magnetic fields were fitted collectively to a Zeeman splitting model,[33–35] with the assumption that the PCS spectra for individual exciton modes near their resonance are Lorentzian.[25] The measured PCS current $I(B, \lambda)$ under magnetic field $B$ and wavelength $\lambda$ is thus described by the following equation:

$$I(B, \lambda) = \frac{\frac{|A_B| \pi \cdot w_B}{2}}{\left[\frac{hc}{\lambda} - \left(X_0^A + \frac{1}{2}\tau \cdot \mu_{Bohr} \cdot g \cdot B\right)\right]^2 + \left(\frac{w_B}{2}\right)^2} \quad (1)$$

Here $h$ is the Planck constant, $c$ is the speed of light, $\mu_{Bohr}$ is the Bohr magneton, $\tau = \pm 1$ is the valley index corresponding to RCP and LCP, and $A_B$ and $w_B$ are the magnetic field-dependent Lorentzian amplitude and linewidth for each spectrum taken, respectively. Equation (1) also includes two magnetic field-independent parameters, where $g$ is the Landé g-factor, and $X_0^A$ is the central energy of the bright A-exciton peak under zero magnetic field. The normalized data (colored solid dots) and model fitting result for each magnetic field is shown as the colored dash line in Fig. 4(b). The valley selection mechanism[20,36] allows a selective excitations of bright intra-valley A-excitons in the K (K′) valley using the RCP (LCP) light, showing a red (blue) shift in the corresponding A-excitonic energy with increasing magnetic fields. This is due to the magnetic field-induced Zeeman splitting in the K and K′-valleys, as illustrated in Fig. 4(c). The fitted result showed that at zero magnetic field, the centers of the $X_0^A$ was found to locate at 1.956 eV, shown as the vertical dark red dashed lines in Fig. 4(b) for both RCP and LCP measurements, whereas the g-factor was found to be $-14.58 \pm 0.61$ and $-11.14 \pm 0.65$ for RCP and LCP, respectively.

Here we remark that the magnitude of the g-factor derived from our PCS studies is significantly larger than that of the g-factor $-4$ derived from photoluminescence measurements,[37–39] which may be attributed to the existence of magnetic field-enhanced formation of intervalley excitons under the presence of an in-plane electric field. The notion of magnetic field-enhanced formation of intervalley excitons can be corroborated by our observation that the linewidth difference, defined as $(w_B^{RCP} - w_B^{LCP})$, was always positive for positive magnetic fields (e.g., 21.5 meV at 12 T) and negative for negative magnetic fields (e.g., $-22.3$ meV at $-12$ T), which implied that positive magnetic fields enhanced (suppressed) the formation of K → K′ intervalley excitons under the excitation of RCP (LCP) light, as manifested by the energy diagrams in Fig. 4(c), which resulted in the broadening (sharpening) of the linewidth $w_B^{RCP}$ ($w_B^{LCP}$). Similarly, negative magnetic fields enhanced (suppressed) the formation of K′ → K intervalley excitons under the excitation of LCP (RCP) light, thus resulting in the broadening (sharpening) of the linewidth $w_B^{LCP}$ ($w_B^{RCP}$). It is worth pointing out that the g-factor values determined from our PCS studies were comparable to the intervalley exciton g-factor of $-12$ reported for monolayer tungsten diselenide.[40] Finally, we note that the slight difference between the g-factors extracted from the RCP and LCP measurements using our PCS may be due to in-plane bias voltage-induced valley symmetry breaking.

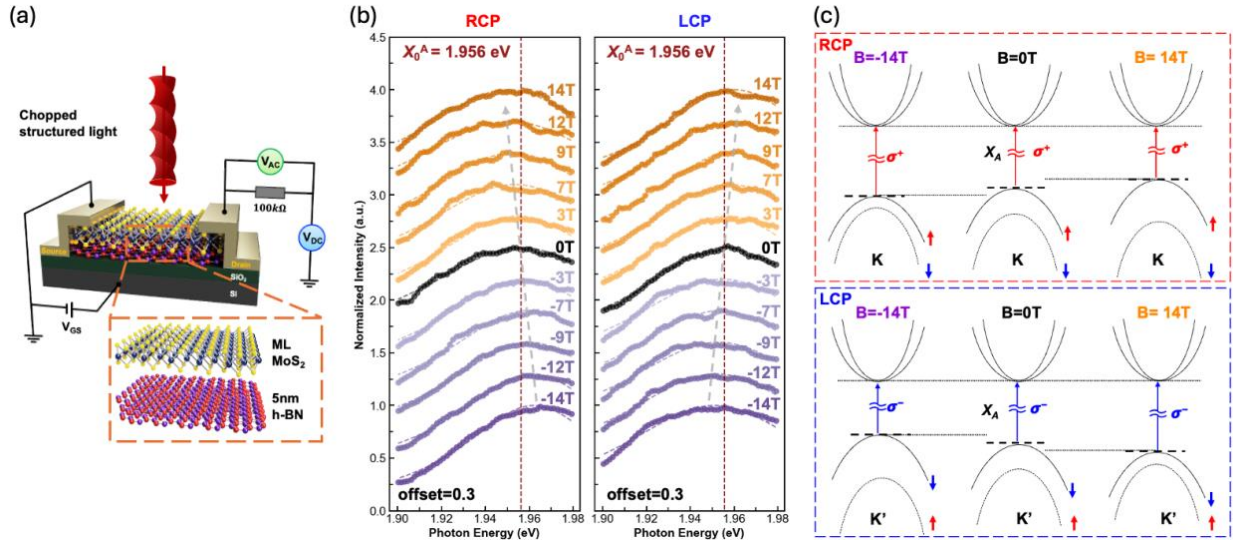

**FIG. 4.** Application of the SAM/OAM-resolved PCS system to deriving the Landé g-factor of ML-MoS$_2$. (a) Schematics of the ML-MoS$_2$ FET device and the electrical circuit for photocurrent measurements. (b) PCS of ML-MoS$_2$ FET between B = 0 and ±14 T under RCP light and LCP light, respectively. The colored dashed lines are fitting results under different magnetic field and the dark red vertical line indicates the fitted center of A-exciton under zero magnetic field, $X_0^A$. The applied $V_{DS}$ is 3 V and $V_{GS}$ is 28V. (c) Energy band schematics of ML-MoS$_2$ under magnetic field with K (red) and K' (blue) valley, respectively. The spin-orbit coupling at the conduction band minimum was small and thus omitted in the schematics for simplicity.

**B. Scanning photocurrent spectroscopy**

The scanning capability of our instrument was evaluated by measuring the spatially dependent RCP-PCS on a ML-MoS$_2$ FET, with approximately twenty photocurrent spectra taken over a spatial region between the source (S) and drain (D) contacts and the sample, as illustrated in the optical micrograph in Fig. 5(a), where the red arrow indicates the scanning direction, and the red dots represent the locations where the spectra were taken. The measured spectra were fitted then normalized to the bright A-exciton peak amplitude as shown in Fig. 5(b). The A-exciton ($X^A$) and B-exciton ($X^B$) peak positions are marked by the red (1.956 eV) and blue (2.098 eV,) dashed lines in Fig. 5(b), respectively, indicating a 142 meV ± 4.3meV splitting of the valence band due to spin-orbit coupling. The extracted amplitudes of $X^A$ and $X^B$ were shown in Fig. 5(c), where amplitude homogeneity in the sample region was apparent. The sharp increase of both $X^A$ and $X^B$ amplitudes near the source contact was due to band-bending-induced carrier saturation, similar to the findings in previous reports.[41,42] Over the metal contact regions, in addition to the expected sharply diminished exciton amplitudes [Fig. 5(c)], the emergence of the A-trion ($T^A$, black dashed line in Fig. 5(b)) accompanied by a suppressed B-exciton amplitude was indicative of exciton-plasmon interactions[43,44] and charge transfering.[45,46]

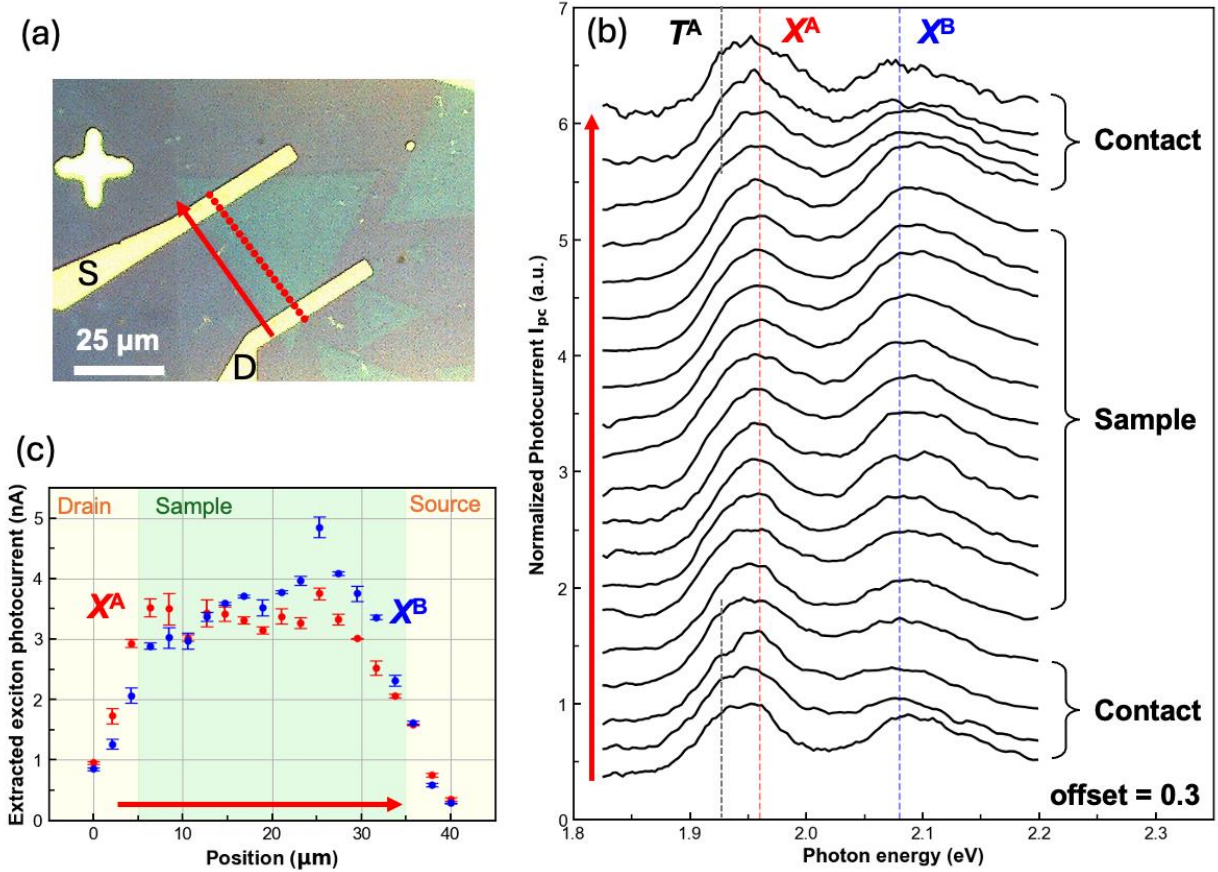

**FIG. 5.** Scanning RCP-PCS on a ML-MoS$_2$ FET. (a) Optical micrograph of the FET device. The red arrow indicates the scanning direction, and the array of red spots marks the measurement locations. (b) Spatially dependent photocurrent spectra normalized to the bright A-exciton peak amplitude. Each spectrum was taken at one of the spots marked in the red dot array in (a). (c) Extracted A and B exciton peak amplitudes from (b). Yellow and green background colors correspond to the regions of metallic contacts and sample, respectively.

### C. OAM-resolved photocurrent spectroscopy

In additional to studies of the SAM-resolved PCS, the effect of OAM light on the PCS of ML-MoS$_2$ was investigated and shown in Fig. 6. We found that the measured photocurrent exhibited a significant increase with increasing $|\ell|$ for both positive [Fig. 6(a)] and negative $\ell$ [Fig. 6(b)], where the controlled $\ell = 0$ spectrum is shown in black for reference in both plots. The extracted peak photocurrents on resonance with the bright A and B-excitons are shown in Fig. 6(c), where up to ~ 80% enhancement in the photocurrent was observed for $\ell = \pm 5$ relative to that with $\ell = 0$. This unusual enhancement was not originated from larger effective spot size under higher $\ell$ because the channel length (~ 30 μm) of the FET device well exceeded the spot size, and moving the spot position over the sample region did not incur discernible changes in the measured photocurrent

[Fig. 5(c)]. Moreover, the laser power variation under different $\ell$ values had been carefully calibrated to be less than ±3% [Fig. 3(d)].

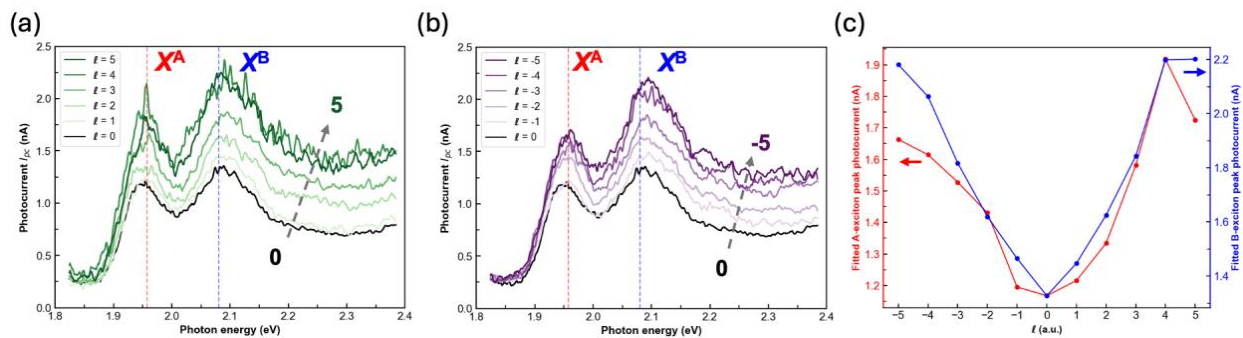

**FIG. 6.** Orbital angular momentum enhanced photocurrent under linearly polarized light. [(a),(b)] PCS obtained under OAM carrying $\ell$ from (a) 0 to 5 and (b) 0 to −5 with linearly polarized light. (c) Fitted A-exciton (red) and B-exciton (blue) peak photocurrents measured at their respective resonance energies under OAM light of different $\ell$.

The substantial increase in photocurrents with increasing $|\ell|$ may be attributed to several important characteristics associated with the OAM light that enables far more channels of excitations in materials relative to light without OAM, which include: rapidly increasing accessible final states in the conduction bands due to increasing momentum transfer from OAM light to electrons[47] with increasing $|\ell|$; unlocking intervalley dark excitonic transitions;[40] enhancing the formation of Rydberg excitons;[48] and enabling quadrupole light-matter interactions in addition to the dipole interaction.[48] Moreover, the rapid increase in the measured photocurrent at ~2.14 eV under light with increasing $|\ell|$ is consistent with strong excitation enhancement of the Rydberg series of the A-exciton reported between 2.089 eV ($A_{2s}$) to 2.172 eV ($A_{5s}$),[25] although individual peaks of the Rydberg excitons could not be resolved from our current PCS measurements due to the limited signal-to-noise ratio and energy resolution, which may be improved in the future by encapsulating the ML-TMD FET devices with layers of h-BN to enhance the light-matter interaction.

**V. Discussion and outlook**

In this work, we reported a new instrumentation development of a cryogenic scanning photocurrent spectroscopy (PCS) system capable of resolving both spin and orbital angular momentum (SAM/OAM) of light for a spectral range from 500 nm (2.48 eV) to 700 nm (1.77 eV) and mapping a (35 × 25) μm² area with ~ 1 μm spatial resolution. We further demonstrated the capabilities of the instrument by applying it to investigate the optoelectronic responses of CVD-grown ML-MoS$_2$ FET devices at 3 K. We found that the SAM-resolved magneto-photocurrent measurement resulted in valley-dependent Zeeman splitting of the bright A-exciton peak, yielding a large Landé *g*-factor possibly due to magnetic field-enhanced intervalley excitons that were assisted by in-plane electric fields. Additionally, the OAM-resolved measurements revealed significant photocurrent enhancement with increasing $|\ell|$, which is consistent with substantially increased excitation channels associated with the OAM light. Further improvements of the signal-to-noise ratio for the PCS measurements may be achieved by encapsulating the ML-TMD FET device with finite thicknesses of hexagonal boron nitride (h-BN) to enhance the light-matter interaction and the magnitude of photocurrents so that the individual energies of dark and Rydberg excitons can be resolved. Additionally, the measurement uncertainty in laser power can be mitigated by using calibrated photodiodes with lock-in techniques to replace the standalone power meter. While there is clearly room for further improvements, the overall design concept of the instrument reported here provides a unique and versatile platform for investigating the optoelectronic responses of diverse quantum materials to structured SAM/OAM light.

**Acknowledgements**

The work at Caltech was supported by the National Science Foundation (NSF) under the Major Research Instrument (MRI) award #DMR-2117094 and the Physics Frontier Center award #1733907 for the Institute for Quantum Information and Matter (IQIM). Authors from the National Taiwan Normal University (NTNU) acknowledge support from the National Science and Technology Council (NSTC) in Taiwan under Contracts NSTC 113–2112-M-003–001, NSTC 112-2926-I-003-504-G, and NSTC 111-2112-M-003-008. C.-I.L. acknowledged support from NSTC in Taiwan under Contracts NSTC 109-2917-I-564 -003 and NSTC 112-2112-M-492-001-MY3. Z. S. and B. R. acknowledged support from Zhiyuan College, Shanghai Jiao Tong University. N.-C.Y. acknowledged support from both the Ministry of Education in Taiwan under the Yushan Fellowship program and NTNU under the Distinguished Yushan Fellow Professorship for sponsoring the collaboration between Caltech and NTNU. Partial support for the device fabrication at the Taiwan Semiconductor Research Institute (TSRI) is also gratefully acknowledged.

**Conflict of Interest Statement**

The authors have no conflicts to disclose.

**Data Availability Statement**

The data that support the findings of this study are available from the corresponding authors upon reasonable request.

## Author contributions

**Duxing Hao**: Conceptualization (equal); Data curation (lead); Methodology (lead); Software (supporting); Investigation (lead); Formal Analysis (lead); Visualization (lead); Writing – Original Draft (lead); Writing – Review & Editing (equal**)**. **Chun-I Lu**: Conceptualization (equal); Resources (supporting); Methodology (supporting); Investigation (supporting). **Ziqi Sun**: Software (supporting); Data curation (supporting); Investigation (supporting); Formal Analysis (supporting); Visualization (supporting). **Yu-Chen Chang**: Conceptualization (supporting); Investigation (supporting); **Wen-Hao Chang**: Investigation (supporting); Writing – Original Draft (supporting); Visualization (supporting). **Ye-Ru Chen**: Resources (supporting); **Akiyoshi Park**: Methodology (supporting). **Beining Rao**: Software (supporting); Formal Analysis (supporting); **Siyuan Qiu**: Software (supporting); Investigation (supporting). **Yann-Wen Lan**: Conceptualization (supporting); Supervision (equal); Resources (supporting); Writing – Review & Editing (supporting). **Ting-Hua Lu**: Conceptualization (equal); Supervision (equal); Validation (equal); Resources (supporting); Writing – Review & Editing (supporting). **Nai-Chang Yeh**: Conceptualization (equal); Investigation (supporting); Methodology (supporting); Supervision (lead); Validation (equal); Resources (lead); Writing – Original Draft (supporting); Writing – Review & Editing (equal).